\documentclass[referee, sn-mathphys-num]{sn-jnl}

\usepackage{amsmath}
\usepackage{graphicx}% Include figure files
\usepackage{bm}% bold math
\usepackage{hyperref}

\usepackage[usenames,dvipsnames,x11names]{xcolor}

\newcommand{\be}{\begin{equation}}
\newcommand{\ee}{\end{equation}}
\newcommand{\bea}{\begin{eqnarray}}
\newcommand{\eea}{\end{eqnarray}}
\newcommand{\ket}[1]{\left\vert #1    \right\rangle }
\newcommand{\bra}[1]{\left\langle   #1  \right\vert}
\newcommand{\ave}[1]{\left\langle #1   \right\rangle }
\newcommand{\prj}[1]{\left\vert #1    \right\rangle \left\langle   #1  \right\vert}

\newcommand{\W}{\Omega}

\begin{document}

\title{Full Qubit Control in the NV$^-$ Ground State for Low Field or High Frequency Sensing}

\author[1]{\fnm{Alberto} \sur{López-García}}

\author*[1]{\fnm{Javier} \sur{Cerrillo}}\email{javier.cerrillo@upct.es}

\affil[1]{\orgdiv{Área de Física Aplicada}, \orgname{Universidad Politécnica de Cartagena}, \orgaddress{ \city{Cartagena} \postcode{30202}, \country{Spain}}}

%\affil[1]{\orgdiv{Área de Física Aplicada}, \orgname{Universidad Politécnica de Cartagena}, \orgaddress{\city{Cartagena} \postcode{30202} \country{Spain}}

\abstract{We present a scheme for the implementation of fast arbitrary qubit gates in the ground state of the negatively charged nitrogen-vacancy (NV$^-$) defect in diamond. The protocol is especially useful in the low-field regime and for high-frequency sensing applications. It constitutes an extension to the NV-ERC technique, which has demonstrated efficient initialization and readout of the double quantum transition with no leakage to any third level thanks to an effective Raman coupling. Here we derive a full theoretical framework of the scheme, identifying the complete unitary associated to the approach, and more specifically the relevant basis change for each of two characteristic pulse durations. Based on this insight, we propose a scheme to perform fast qubit transformations in the double quantum transition. We study its robustness with respect to pulse-timing errors resulting from faulty identification of system parameters or phase-control limitations. We finally demonstrate that the technique can also be implemented in the presence of unknown electric or strain fields.
}


\maketitle

\section{Introduction}

Quantum sensing with negatively charged nitrogen-vacancy (NV) centers in diamond \cite{Jelezko2002,Jelezko2004,Doherty2013} has become an established tool in the fields of magnetometry \cite{Staudacher2013,Mamin2013,Muller2014,Ajoi2015,Lovchinsky2016,Glenn2018}, thermometry or electrometry \cite{Dolde2011,Neumann2013}. Its ground state is a spin-1 triplet that can sustain coherence for long times even at ambient conditions \cite{Childress2006,Balasubramanian2008,Maze2008,Jacques2009}, supporting its application as a nuclear magnetic resonance (NMR) spectrometer at the nanoscale \cite{Lukin2012,Hanson2012,Zhao2012,Laraoui2013}. This goal requires the development of control strategies that are specifically tailored to the features of the system.

Nevertheless, even though the ground state of NV centers is a spin-1 triplet, conventional operation ignores the highest lying state \cite{Maudsley1986,Souza2011,Wang2011,Kotler2011,Souza2012, Cywinski2007,Uhrig2008,Uhrig2007a,deLange2010,Ryan2010,Naydenov2011,Casanova2015,Abobeih2018,Wang2019}. This approach requires that the spectral separation between the sensing transition and the third state is larger than the strength of the pulses. Away from this regime, leakage to the third level prevents operation \cite{London2014} and results in complementary limitations in sensing: on the one hand, high frequency sensing becomes impossible with typical dynamical decoupling approaches and, on the other hand, low field operation becomes unfeasible.

The NV effective Raman coupling (NV-ERC) approach \cite{Cerrillo2021} solves this fundamental limitation through the design of pulses that take full account of the coupling to the third level. By driving with a frequency that corresponds to the zero-field splitting,
%in the middle point of the double quantum transition,
it is possible to initialize the NV center to the equator of the Bloch sphere of the double quantum transition without loss to a third state. This approach has been experimentally applied  \cite{Vetter2022,Li2024}, where attempts were made at exploiting the phase of pulses to perform effective $\pi$ rotations in the double quantum transition. This ability is crucial for the implementation of complex dynamical decoupling protocols. Nevertheless, lack of a full theoretical picture of the mechanism behind NV-ERC has so far prevented the design of elaborate operations for arbitrary values of magnetic field and microwave power and precluded the study of the limitations and effects of external perturbations, such as electric or strain fields. Here we resolve these shortcomings by means of a complete theoretical analysis of NV-ERC and a proposal for arbitrary qubit gate implementation.
The regime of low Zeeman splitting is essential for ultra-low field detection or environment quantum control \cite{Reinhard2012,Laraoui2013,Walsworth2018,Schmitt2017,Ajoi2019,Zheng2019,Walsworth2020}. In the zero-field limit the effective Raman coupling vanishes but coupling to the symmetric superposition of degenerate states $\ket{-1}$ and $\ket{+1}$ remains. This has been exploited in previous zero-field works \cite{Sekiguchi2016,Saijo2018,Toily2013,Hodges2013,Fang2013,Sekiguchi2019,Sekiguchi2017} but none can be directly applied to the finite field case discussed here.

We first fully identify the evolution unitary operator for any pulse duration in section \ref{sec:Fundamentals} and proceed in section \ref{sec:Pulses} to derive the full basis transformation associated with the two characteristic times of state preparation and readout. We show in sec.\ref{sec:Rotations} that these two transformations may be combined to create arbitrary fast single qubit gates. As a discussion of the presented results, we explore the robustness of the approach in sec.\ref{sec:Robustness} with respect to errors in the identification of experimental parameters, pulse duration or control inaccuracies. Finally, in section \ref{sec:Strain} we show that the effect of strain and/or electric field perturbation does not prevent the implementation of the scheme.

\section{Results}
\subsection{NV-ERC Fundamentals}\label{sec:Fundamentals}
We consider the ground state of an NV center with quadrupolar splitting $D$ subject to the effect of a magnetic field $B$ aligned with its $z$ axis and driven by a MW pulse of frequency $D$ and amplitude $\W$ as expressed by the Hamiltonian
\be
H=DS_z^2 + \mu B S_z + \W \cos \left( Dt - \alpha \right) S_x,
\label{eq:H}
\ee
where we allow for an arbitrary phase $\alpha$ associated with the driving. Based on the $S_z$ eigenstates $\ket{0}$, $\ket{+1}$ and $\ket{-1}$, it is useful to define the states
\bea
\ket{+}&=&\frac{1}{\sqrt{2}}\left(\ket{+1}+\ket{-1}\right),\\
\ket{-}&=&\frac{1}{\sqrt{2}}\left(\ket{+1}-\ket{-1}\right),
\eea
so that, by moving into the interaction picture with respect to $H_0=DS_z^2$ and taking the rotating wave approximation, we end up with the expression
\be
H'=\mu B \ket{-}\bra{+} +\frac{\W}{2} e^{i \alpha} \ket{+} \bra{0} + H.c.,
\label{eq:HRWA}
\ee
corresponding to an effective Raman coupling between states $\ket 0$ and $\ket{-}$. Resonant Raman couplings have a straightforward analytical solution through the definition of a bright and a dark state
\bea
\ket{B_\alpha}&=&\frac{1}{\bar\W}\left(\mu B \ket{-} + e^{-i\alpha}\frac{\W}{2}\ket{0}\right),\\
\ket{D_\alpha}&=&\frac{1}{\bar\W}\left(-\mu B \ket{0} + e^{i\alpha}\frac{\W}{2}\ket{-}\right),
\eea
with $\bar\W^2=\mu^2 B^2 + \W^2/4$. The dependence on the pulse phase $\alpha$ is made explicit as it plays a central role in the definition of arbitrary gates in the double quantum subspace. The evolution unitary operator associated with Hamiltonian eq.(\ref{eq:HRWA}) is finally
\be
\label{eq:U}
U(t, \alpha)=\cos \bar\W t\left(\prj{B_\alpha}+\prj{+}\right)-i\sin\bar\W t\left(\ket{B_\alpha}\bra{+}+\ket{+}\bra{B_\alpha}\right)+\prj{D_\alpha},
\ee
representing a Rabi oscillation between states $\ket +$ and $\ket{B_\alpha}$ of period $\bar T/2=\pi/\bar\W$.

\subsection{Beyond NV-ERC}\label{sec:Pulses}
NV-ERC is based on a unique feature of $U(t, \alpha)$ by which there exists a time $\bar T'$ that maps all population from the ground state $\ket 0$ to the equator of the double quantum Bloch sphere defined by states $\ket{+1}$ and  $\ket{-1}$. This may be expressed $\bra{0}U(\bar T', \alpha)\ket{0}=0$, which from eq.(\ref{eq:U}) implies that $\bar T'$ must satisfy
\be
\cos\bar\W\bar T' \left|\ave{0|B_\alpha} \right|^2+\left|\ave{0|D_\alpha} \right|^2=0
\label{eq:cond}
\ee
and therefore
\be
\bar T'=\frac{\arccos\left(-\frac{4\mu^2B^2}{\W^2}\right)}{\bar\W},
\ee
which is independent of $\alpha$. After a pulse of this length, the state $\ket{0}$ is mapped to state $\ket\phi$ as defined by 
\be
\ket\phi=e^{-i\frac{\phi}{2}}\ket{-1}-e^{i\frac{\phi}{2}}\ket{+1}
\ee
with $\phi=\arccos(2\mu B/\W)$. This mapping is independent of phase $\alpha$ and can be implemented for any value of $\W\geq2\mu B$, that is, in the fast pulse or low field regimes.

\begin{figure}
    \centering
    \includegraphics[width=0.7\columnwidth,trim={ 4.8cm 4.4cm 5.5cm 2.2cm},clip]{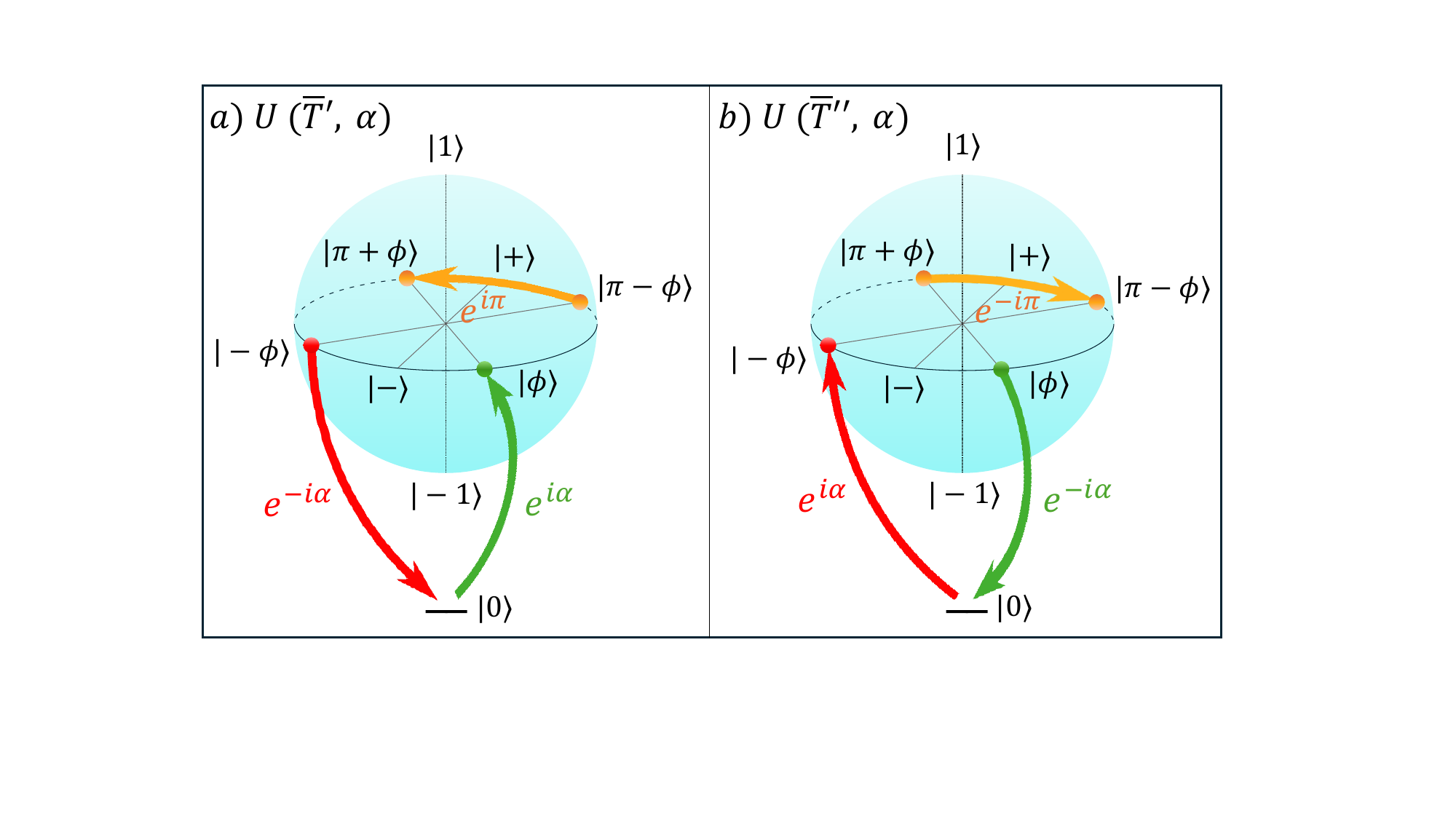}
    \caption{Illustration of the effect of the unitary transformations (a) $U(\bar T', \alpha)$ eq.(\ref{eq:Ut1}) and (b) $U(\bar T'', \alpha)$ eq.(\ref{eq:Ut2}). Due to the relevance of coherences between states $\ket{-1}$ and $\ket{+1}$, their corresponding Bloch sphere is depicted. This allows to represent the four states $\ket{\phi}$, $\ket{-\phi}$, $\ket{\pi-\phi}$ and $\ket{\pi+\phi}$ that are involved in both unitaries. Each of the three simultaneous mappings are indicated by an arrow of a different colour, together with the phase gain associated with it.}
    \label{fig:Ut}
\end{figure}

In this paper we explore this feature in the full Hilbert space by finding the complete unitary in terms of the parametrization $\phi$. As shown in the appendix, we may write $U(\bar T', \alpha)$ as a basis transformation
\be
U(\bar T', \alpha)=e^{i\alpha}\ket{\phi}\bra{0}+e^{-i\alpha}\ket{0}\bra{-\phi}-\ket{\pi+\phi}\bra{\pi-\phi},
\label{eq:Ut1}
\ee
which provides us with a deeper intuition of the behaviour of the three level system for pulses of duration $\bar T'$ as well as the effect of the phase $\alpha$. First of all, the  mapping of $\ket 0$ into $\ket \phi$ comes with the acquisition of a phase $\alpha$. In addition, the state $\ket{\pi+\phi}$, which is orthogonal to $\ket \phi$ originates in state $\ket{\pi-\phi}$, , i.e. it can be seen as the result of a $2\phi$ rotation of the double quantum Bloch sphere. This rotation involves a phase gain of $\pi$ regardless of the value of $\alpha$. Finally, the state $\ket{-\phi}$, that is orthogonal to $\ket{\pi-\phi}$, is mapped to $\ket{0}$ with a phase gain of $-\alpha$. All in all, as illustrated in fig.\ref{fig:Ut}(a), the full unitary amounts to the orthonormal basis transformation
\be
\{\ket 0, \ket{\pi-\phi}, \ket{-\phi}\}\xrightarrow{\bar T'}\{e^{i\alpha}\ket \phi, -\ket{\pi+\phi}, e^{-i\alpha}\ket{0}\}.
\ee

A similar interpretation may be gained for a pulse of duration $\bar T''$. By noting that the unitary for $\bar T=2\pi/\bar\W$ is the identity and that $\bar T''=\bar T-\bar T'$, it must perform the opposite basis mapping
\be
\{\ket 0, \ket{\pi-\phi}, \ket{-\phi}\}\xleftarrow{\bar T''}\{e^{i\alpha}\ket \phi, -\ket{\pi+\phi}, e^{-i\alpha}\ket{0}\},
\ee
and we may write
\be
U(\bar T'', \alpha)=e^{i\alpha}\ket{-\phi}\bra{0}+e^{-i\alpha}\ket{0}\bra{\phi}-\ket{\pi-\phi}\bra{\pi+\phi},
\label{eq:Ut2}
\ee
as illustrated in fig.\ref{fig:Ut}(b).

\begin{figure}
    \centering
    \includegraphics[width=0.7\columnwidth]{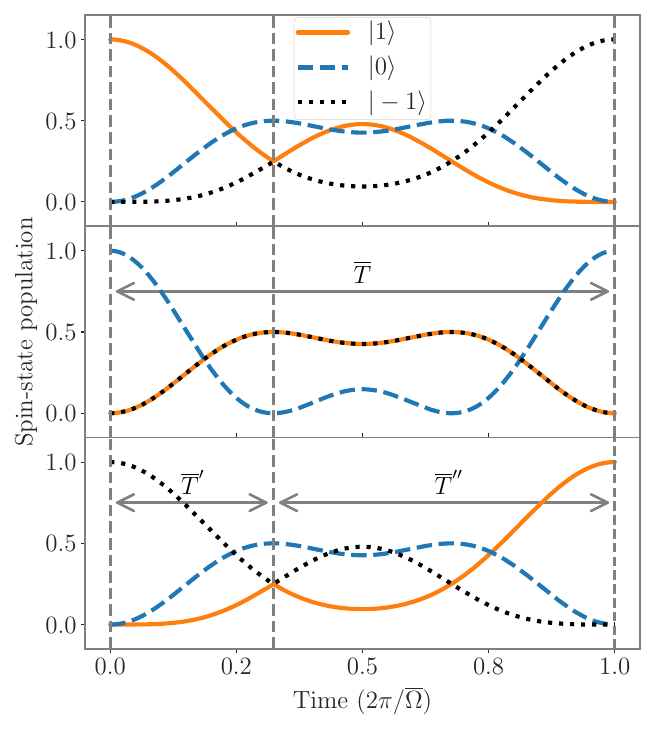}
    \caption{Time evolution of the populations of the $|1\rangle$, $|0\rangle$ and $|-1\rangle$ states during the implmentation of a NOT gate in the double quantum transition consisting of a first pulse of duration $\bar{T}'$ with $\Omega = 3\mu B$ and a second pulse of duration $\bar{T}''$ with $\Omega = -3\mu B$. The vertical dashed gray lines represent the time periods $\bar{T}'$ and $\bar{T}''$,  which together constitute the total time period $\bar{T}$.}
    \label{fig:PiPulse}
\end{figure}

\subsection{Arbitrary Rotations}\label{sec:Rotations}

One may use the previous insight to design pulse combinations that perform transformations of interest. Here we show that it is sufficient to consider the concatenation of two pulses to perform tailored arbitrary rotations in the double quantum Bloch sphere corresponding to the levels $\ket{-1}$ and $\ket{+1}$.

Let us first consider a pulse of duration $\bar T'$ and phase $0$, followed by another one of duration $\bar T''$ and phase $\pi$, resulting in a concatenation of total duration $\bar T= \bar T'+\bar T''$. Throughout, the Rabi frequency $\W$ and magnetic field $B$ are kept constant. As illustrated in figure \ref{fig:PiPulse}, this amounts to a NOT gate operated between states $\ket{+1}$ and $\ket{-1}$ that does not affect $\ket{0}$. This is close to the result found in \cite{Vetter2022} by means of optimal control theory and that constituted an improved version of the $\pi$ pulse of duration $\bar T/2$ proposed in \cite{Cerrillo2021} for the regime $\W\gg \mu B$. Pulses of this form permit implementation of dynamical decoupling sequences as was shown in \cite{Vetter2022}.

Now we can use the derivations of the previous section to fully capture the mechanism behind this $\pi$ pulse and propose generalizations thereof. Let us consider the more general case of a first pulse of duration $\bar T'$ with phase $\alpha$ and a second one of duration $\bar T''$ and phase $\alpha+\theta$. By combining eq. \ref{eq:Ut1} and eq. \ref{eq:Ut2}, one obtains
\bea
R(-\phi,\theta)&=&U(\bar T'', \alpha +\theta)U(\bar T', \alpha)\\
\nonumber &=&e^{-i\theta}\ket{0}\bra{0}+e^{i\theta}\ket{-\phi}\bra{-\phi}+\ket{\pi-\phi}\bra{\pi-\phi},
\eea
which amounts to a rotation of angle $\theta$ around the axis $-\phi$. The case $\theta=\pi$ corresponds to fig. \ref{fig:PiPulse}, and clarifies two points. First, the absolute value of the phases of each pulse is not relevant, but rather the difference between their phases. Second, we have now identified the specific axis around which the NOT gate is performed. Interestingly, this axis will depend on the ratio between $\W$ and $\mu B$. This constitutes a first means to achieve control over the axis around which the rotations are performed. In particular, $\pi$ pulses of different intensities could suffice to implement common DD sequences such as XY8.

\begin{figure}
    \centering
    \includegraphics[width=.4\columnwidth,trim={ 10cm 6.1cm 13cm 4.2cm},clip]{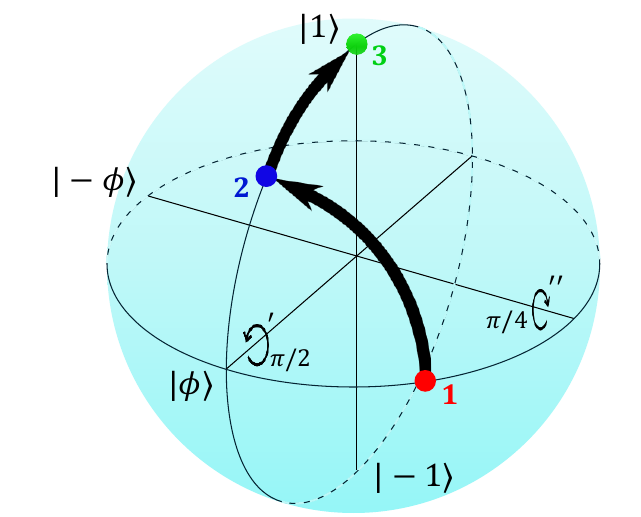}
    \caption{Demonstration of directed manipulation and positioning capability of an initial quantum state through its dynamic unitary evolution for applications in dynamical decoupling . The initial state, marked with the red dot (1), is in an intermediate position between the $|\phi\rangle$ and $|\pi - \phi\rangle$ states. A rotation pulse $\pi/2$ is then applied along the $|\phi\rangle \leftrightarrow |\pi + \phi\rangle$ axis, which transforms the initial state (1) to the state marked with the blue dot (2). Finally, a $\pi/4$ rotation pulse is applied along the $|-\phi\rangle \leftrightarrow |\pi - \phi\rangle$ axis, which transforms state 2 to the final state marked in green (3), corresponding to state $|1\rangle$.}
    \label{fig:Gate}
\end{figure}

A second alternative does not require adjusting the intensity of the pulses. It is achieved by simply inverting the order of the pulse concatenation, which results in a rotation around the axis $\phi$ instead of $-\phi$
\bea
R(\phi,\theta)&=&U(\bar T', \alpha+\theta)U(\bar T'', \alpha)\\
\nonumber&=&e^{-i\theta}\ket{0}\bra{0}+e^{i\theta}\ket{\phi}\bra{\phi}+\ket{\pi+\phi}\bra{\pi+\phi}.
\eea
As long as $\W>2\mu B$, $\phi>0$ and both rotations will occur on different axes. It is a known result that any gate can be decomposed in up to three rotations around two non-parallel axes, in our case $R(\pm \phi, \theta)$. We illustrate this with the diagram fig.\ref{fig:Gate}, where two rotations are used to polarize the NV center along state $\ket{+1}$. This establishes the present scheme as a means to achieve arbitrary rotations in the Bloch sphere with the concatenation of pulses for a total duration of, at most, $3\bar T$.

\section{Discussion}
\subsection{Robustness}\label{sec:Robustness}

\begin{figure*}
    \centering
    \includegraphics[width=\textwidth]{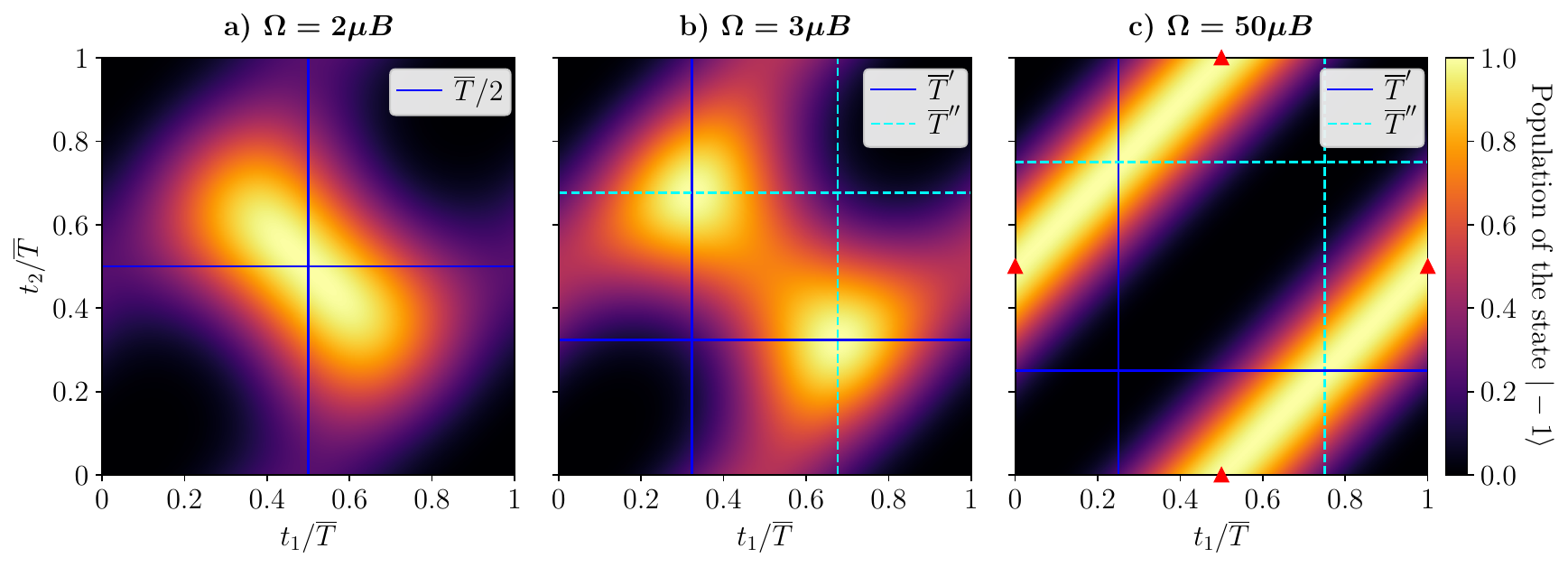}
    \caption{Population density map of the $|-1\rangle$ state as a function of pulse times $\bar{T}$, $\bar{T}'$ and $\bar{T}''$, under different settings of the $\Omega$ parameter. The superimposed lines — $\bar{T}/2$ in (a) (solid blue), $\bar{T}'$ and $\bar{T}''$ in (b) and (c) (solid blue and dashed light blue, respectively) — indicate the characteristic pulse times that produce critical changes in the system population. In (c), the red triangle markers refer to critical pulse times identified in previous analysis \cite{Cerrillo2021}.}
    \label{fig:Robustness}
\end{figure*}

In this section we consider the effect of static errors in pulse implementation. We can consider different sources for the errors: (a) experimental parameters, (b) phase change sluggishness and (c) timing. We argue that the first two can be reinterpreted as pulse timing errors.

Experimental parameters such as Rabi frequency $\W$ or Zeeman splitting $\mu B$ may vary slightly between characterization and experiment realization. The effect this would have is a wrongly identified $\bar\W$ and therefore an erroneous value of $\bar T$, resulting in total pulse duration above or below $\bar T$. Additionally, unless the error in $\W$ and $\mu B$ is proportional, this would also lead to an error in $\bar T'$.

Phase change sluggishness can be modeled by a modification of the phase $\alpha$ away from the intended $\bar T'$ or $\bar T''$. Without errors in the experimental parameters, the concatenation of both pulses would still correctly match the total duration $\bar T$ but with incorrectly assigned times $\bar T'$ and $\bar T''$ to the phases $\alpha$ and $\alpha +\theta$.

After identifying that two separate error sources can be approximately modeled as timing errors, we analyze the ability of the pulse sequence to  produce a $\ket{+1}\rightarrow\ket{-1}$ transformation with the pulse combination $U(t_2,\pi)U(t_1,0)$ and $\theta=\pi$. This is analyzed in fig.(\ref{fig:Robustness}) for three different values of the ratio $\W/\mu B$.

For the minimum ratio $\W/\mu B=2$, the pulse combination produces a $\pi$ pulse in the case $t_1=t_2=\bar T /2$. This pulse is more sensitive to the coordinate $t_1+t_2$, that is, an error in the total duration of the pulse, whereas it is less sensitive to an error in the splitting of the pulses. Following our previous discussion, phase sluggishness in this regime is not too relevant as long as the total time $\bar T$ is accurately known. The other limit, where $\W/\mu B=50$, the opposite effect takes place. Two timing combinations produce the expected transformation, either $t_1=\bar T'$ and $t_2=\bar T''$, or $t_1=\bar T''$ and $t_2=\bar T'$ . Additionally, it is possible to identify four other locations where an approximate $\pi$ pulse takes place as identified in \cite{Cerrillo2021}. This corresponds to a situation when one of the pulses has no effect and the other one has a duration of $\bar T/2$. This makes this combination especially insensitive to total pulse duration $t_1+t_2$ as long as their difference matches $|t_1-t_2|=\bar T''-\bar T'\simeq\bar T/2$. As discussed earlier, this regime is especially useful in the case where the phase timing can be well controlled but there might exist a misidentification of $\W$ or $\mu B$.

\subsection{Effect of Strain and Electric Fields}\label{sec:Strain}

We now turn to consider the effect of strain and electric fields in the scheme described so far. These can be introduced as an additional potential to the Hamiltonian $H$ in eq.(\ref{eq:H})
\be
V=E_zS_z^2+E_x\left(S_y^2-S_x^2\right)+E_y\left(S_xS_y+S_yS_x\right),
\ee
where we lump the effect of the electric field and strain components into the energy terms $E_j$ with $j\in\{x,y,z\}$.
Assuming the presence of unknown static fields, we argue their effects can be circumvented in an implementation of the protocol. Experimentally, all required parameters can be obtained in a two-step process:
\begin{enumerate}
\item An initial ODMR process is performed to identify the energies of the three ground states of the NV center, in order to first identify the frequency of the pulses and correctly align the magnetic field. 
\item A Rabi experiment is performed to identify which pulse length transfers all population away from state $\ket 0$, which corresponds to $\bar T'$ and the total cycle time $\bar T$.
\end{enumerate}
We show below that this procedure automatically captures the effect of most perturbations introduced by static electric or strain fields.

\subsubsection{Fields in the $z$ direction}

The effect of $E_z$ is rather innocuous to the scheme, since it only shifts the value of $D$ to $D+E_z$. This shift would be picked up in the ODMR step, so pulses would automatically be designed with the appropriate frequency $D+E_z$. In order to better understand whether the effect of the remaining components of electric and strain fields can be detrimental to the scheme, it is convenient to incorporate their effect in the RWA Hamiltonian $H'$ in eq.(\ref{eq:HRWA}), this time calculated with respect to the interaction picture $(D+E_z)S_z^2$
\be
\label{eq:HX}
H''=E_x\left(\prj{+}-\prj{-}\right)+\left[\left(\mu B + i E_y\right) \ket{-}\bra{+} +\frac{\W}{2} e^{i \alpha} \ket{+} \bra{0} + H.c.\right].
\ee
We will now consider the effect of each of the two remaining components.

\begin{figure}
    \centering
    \includegraphics[width=0.7\columnwidth]{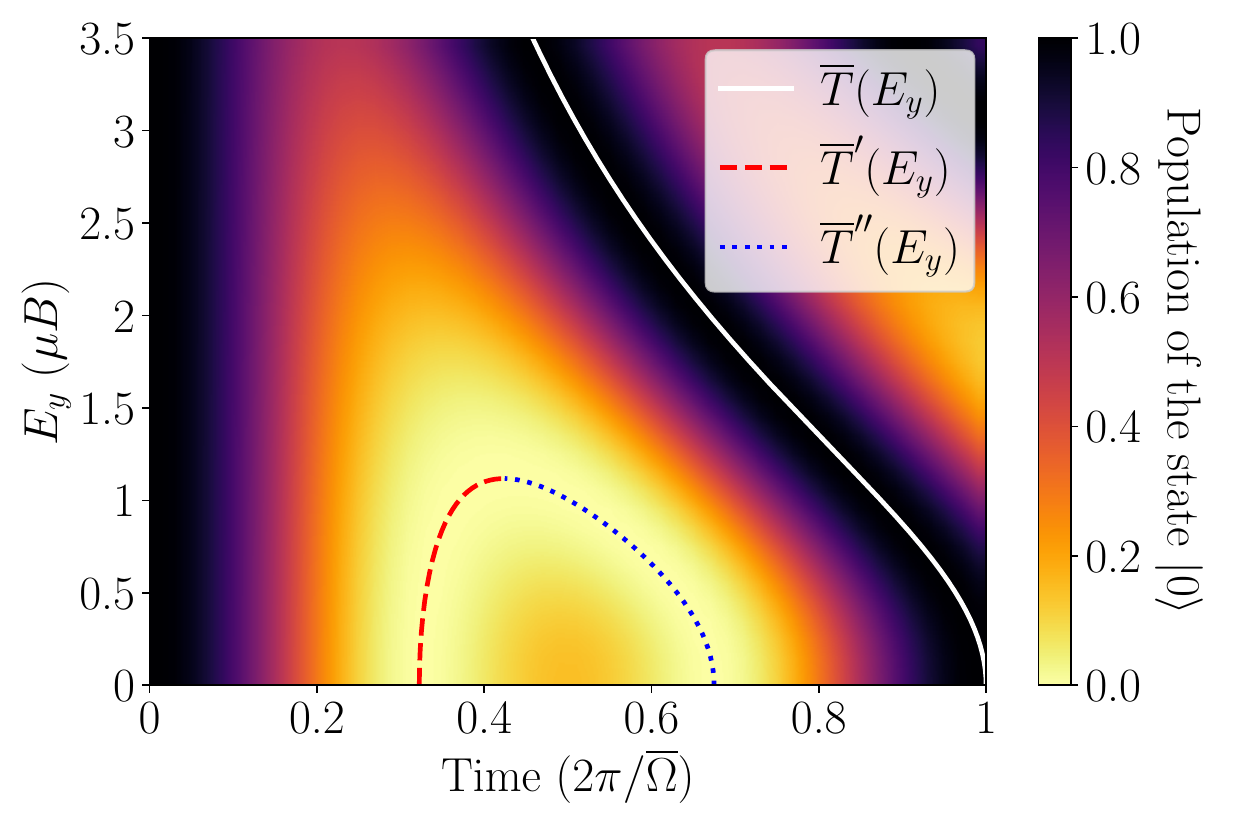}
    \caption{Population density map of the state $|0\rangle$ as a function of electric field $E_y$ and pulse time $T$. The overlapping curves represent the three times: $\bar{T}(E_y)$ (solid white), $\bar{T}'(E_y)$ (dashed red) and $\bar{T}''(E_y)$ (dotted blue) according to eq.(\ref{eq:TEy}) and eq.(\ref{eq:TprimeEy}) with $\Omega = 3\mu B$. For $E_y>\sqrt{5}\mu B/2$, a larger value of $\W$ would be required for the implementation of the protocol.}
    \label{fig:Ey}
\end{figure}

\begin{figure*}
    \centering
    \includegraphics[width=\textwidth]{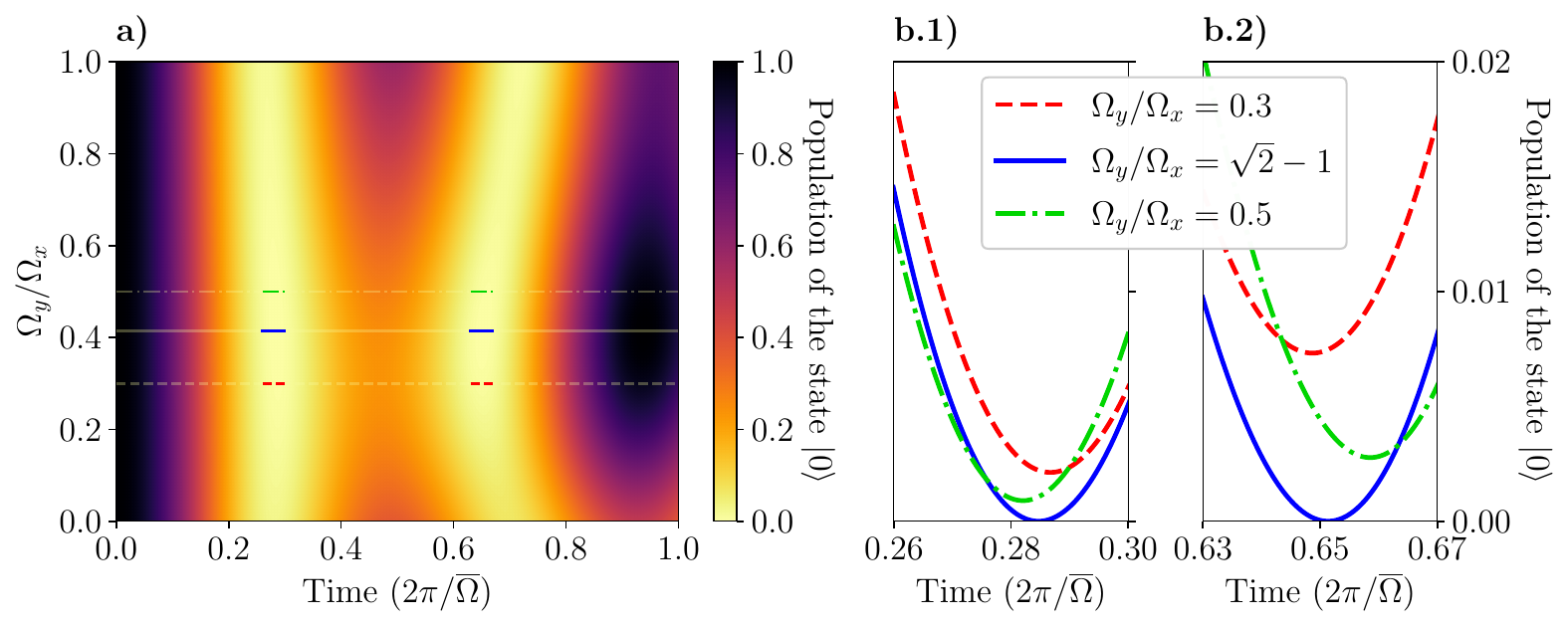}
    \caption{Population density map of the state $|0\rangle$ as a function of the ratio of $\Omega_y$ to $\Omega_x$ (with $\Omega_x = 4.5\mu B$) and the pulse time $T$ in the continuous presence of the electric field $E_y = -E_x = 0.7 \mu B$. The overlapping lines highlighted in specific ratios — 0.3 (dashed red), $\sqrt{2} - 1$ (solid blue) and 0.5 (dashed green) — represent the analysis detailed in b.1) and b.2), which delve into the critical time intervals of the state.}
    \label{fig:Ex}
\end{figure*}

\subsubsection{Fields in the $y$ direction}

Let us first consider the case $E_x=0$, which coincides with $H'$ in eq.(\ref{eq:HRWA}) after the substitutions
\bea
 \mu B&\rightarrow& \sqrt{(\mu B)^2 + E^2_y},\\
 \ket{-}&\rightarrow&\frac{\mu B + i E_y}{\sqrt{(\mu B)^2 + E^2_y}}\ket{-}.
 \eea
The first just amounts to an enhancement of the effect of the magnetic field, whereas the second is a unitary rotation of the double-quantum Bloch sphere around the $x$ axis by an angle $\arctan(E_y/\mu B)$.

The effective enhancement of the magnetic field shortens the total Rabi period $\bar T$
\be
\bar T(E_y)=\frac{2\pi}{\sqrt{\bar\W^2 +E_y^2}},
\label{eq:TEy}
\ee
and modifies the value of $\bar T'$
\be
\bar T'(E_y)=\frac{\arccos \left(-4\frac{\mu^2B^2+E_y^2}{\W^2}\right)}{\sqrt{\bar\W^2 +E_y^2}},
\label{eq:TprimeEy}
\ee
which requires the pulse strength $\W\geq2\sqrt{\mu^2B^2+E_y^2}$. As a result, the value $\bar T''(E_y)=\bar T(E_y)-\bar T'(E_y)$ is also affected. We illustrate these dependencies in fig.\ref{fig:Ey}. Additionally, the state $\ket{\phi}$ is also affected. It is possible to relate the value of $\phi$ to the fraction $\bar T'/\bar T$ of both experimentally obtained times with
\be
\phi=\arccos\left[\sqrt{-\cos\left(2\pi\frac{\bar T'}{\bar T}\right)}\right].
\ee
Summing up, the Rabi experiment to learn $\bar T$ and $\bar T'$ will pick up all these effects, provided sufficient strength of the MW pulse is available. It would still be possible to learn $E_y$ with weak pulse measurements if required by a specific application.

\subsubsection{Fields in the $x$ direction}

The presence of $E_x$ is more delicate since it affects the Raman resonance. Nevertheless, the resonance may be recovered by the addition of an orthogonal microwave in the $y$ direction. We will distinguish the effect of the original MW in the $x$ direction and the additional one in the $y$ direction with Rabi frequency symbols $\W_x$ and $\W_y$ respectively. This produces the Hamiltonian
\bea
\label{eq:HXY}
H'''&=&E_x\left(\prj{+}-\prj{-}\right)+\left[\left(\mu B + i E_y\right) \ket{-}\bra{+} + H.c.\right]\\
&+&\left[\left(\frac{\W_x}{2} e^{i \alpha} \ket{+}+i\frac{\W_y}{2} e^{i \beta} \ket{-}\right) \bra{0} + H.c.\right].\nonumber
\eea
The recovery of the Raman resonance requires adjusting $\beta=\alpha$ and the ratio $\W_y/\W_x$ to the value
\be
\frac{\W_y}{\W_x}=\frac{E_y}{E_x}-\mathrm{sign}\left(\frac{E_y}{E_x}\right)\sqrt{\frac{E_y^2}{E_x^2}+1},
\label{eq:Ratio}
\ee
under which the Hamiltonian eq.(\ref{eq:HXY}) may be rewritten
\be
\label{eq:HXYrot}
\bar H'''=\left(\mu B + i \sqrt{E_y^2+E_x^2}\right) \ket{-}\bra{+} +\frac{\sqrt{\W_x^2+\W_y^2}}{2} e^{i \alpha} \ket{+} \bra{0} + H.c..
\ee
This Hamiltonian is again of the same form as eq.(\ref{eq:HRWA}) under substitutions
\bea
 \mu B&\rightarrow& \sqrt{(\mu B)^2 + E^2_y + E^2_x},\\
 \ket{-}&\rightarrow&\frac{\mu B + i \sqrt{ E^2_y + E^2_x}}{\sqrt{(\mu B)^2 + E^2_y+ E^2_x}}\ket{-}.
 \eea
from which the generalization discussed in the previous subsection applies analogously.

Experimentally, the appropriate ratio eq.(\ref{eq:Ratio}) can be found by testing which one best depletes the ground state $\ket{0}$. In fig.\ref{fig:Ex} we illustrate this approach in the case $E_y=-E_x$, where eq.(\ref{eq:Ratio}) provides the ratio
 $\W_y/\W_x=\sqrt{2}-1$. The two times where the population of the ground state is fully depleted as a function of the ratio $\W_y/\W_x$ correspond to the characteristic times $\bar T'$ and $\bar T''$.

\section{Conclusions\label{sec:Con}}

We propose a scheme for arbitrary qubit gates in the ground state of the negatively charged NV center in diamond that is based on MW pulses tuned to the zero-field transition. This is an extension of the NV-ERC scheme for initialization and readout and its implementation only requires phase modifications.

After analytically finding the evolution unitary operator for the NV-ERC pulse, the basis transformation corresponding to its characteristic pulse durations is derived. This provides a clear insight for the combination of pulses in such a way that arbitrary qubit gates can be performed in the qubit formed by states $\ket{-1}$ and $\ket{+1}$. The robustness of this approach concerning timing errors is explored. We analyzed the effect of unknown static electric and strain fields may have and concluded they can all be well identified and incorporated into the analysis.

\section{Methods}

\subsection{Calculation of the Unitary}

After a pulse of this length, the state $\ket{0}$ is mapped to state $\ket\phi$ as defined by 
\bea
\nonumber &U&(\bar T', \alpha)\ket{0}=e^{i\alpha}\ket{\phi}\\
\nonumber&=&\ave{B_\alpha | 0}\left(\cos \bar\W \bar T' \ket{B_\alpha} -i\sin\bar\W\bar T' \ket{+}\right)+\ave{D_\alpha | 0}\ket{D_\alpha}\\
\nonumber&=&\ave{B_\alpha | 0}\left(\ave{-|B_\alpha }\cos \bar\W \bar T' \ket{-} -i\sin\bar\W\bar T' \ket{+}\right)+\ave{D_\alpha | 0}\ave{-|D_\alpha }\ket{-}\\
\nonumber&=&e^{i\alpha}\left[\frac{\mu B \W}{2\bar\W^2}\left(\cos \bar\W \bar T' -1\right)\ket{-} -i\frac{\W}{2\bar\W}\sin\bar\W\bar T' \ket{+}\right].
\eea
We can further simplify $\ket{\phi}$ by using eq.(\ref{eq:cond}) to find
\be
\nonumber
\cos \bar\W \bar T' -1=-\left|\ave{0|B_\alpha} \right|^{-2},
\ee
and
\be
\nonumber
\left|\ave{0|B_\alpha} \right|\sin \bar\W \bar T' =\sqrt{\frac{1-\cos^2 \bar\W \bar T'}{1-\cos \bar\W \bar T'}}=\sqrt{1+\cos \bar\W \bar T'},
\ee
so finally
\bea
\nonumber\ket{\phi}&=&-\frac{2\mu B}{\W}\ket{-} -i\sqrt{1-\frac{4\mu^2B^2}{\W^2}}\ket{+}\\
\nonumber&=&-\cos \frac{\phi}{2} \ket{-} -i\sin\frac{\phi}{2} \ket{+}\\
\nonumber&=&-e^{i\frac{\phi}{2}} \ket{+1} +e^{-i\frac{\phi}{2}} \ket{-1}.
\eea

The next step is to find which state is mapped into $\ket{0}$ by computing
\bea
\nonumber& \bra{0}&U(\bar T', \alpha)=\ave{0|D_\alpha }\bra{D_\alpha}+\ave{0|B_\alpha}\left(\cos \bar\W \bar T' \bra{B_\alpha} -i\sin\bar\W\bar T' \bra{+}\right)\\
\nonumber&=&\ave{0|D_\alpha}\ave{D_\alpha |-}\bra{-}+\ave{0|B_\alpha }\left(\ave{B_\alpha |-}\cos \bar\W \bar T' \bra{-} -i\sin\bar\W\bar T' \bra{+}\right)\\
\nonumber&=&e^{-i\alpha}\left[\frac{\mu B \W}{2\bar\W^2}\left(\cos \bar\W \bar T' -1\right)\bra{-} -i\frac{\W}{2\bar\W}\sin\bar\W\bar T' \bra{+}\right]\\
\nonumber&=&e^{-i\alpha}\bra{-\phi},
\eea
which implies
\be
\nonumber 
U(\bar T', \alpha)\ket{-\phi}=e^{-i\alpha}\ket{0}.
\ee

Finally, the unitary can be completed by finding the effect of $U(\bar T', \alpha)$ on the state that is orthogonal to both $\ket{0}$ and $\ket{-\phi}$, i.e. $\ket{\pi-\phi}$. This state must be mapped to the one that is orthogonal to both $\ket{0}$ and $\ket{\phi}$, i.e. $\ket{\pi+\phi}$, so $|\bra{\pi+\phi} U(\bar T', \alpha)\ket{\pi-\phi}|=1$. In order to obtain the phase, we compute
\bea
\nonumber&& \bra{\pi+\phi} U(\bar T', \alpha)\ket{\pi-\phi}=\cos^2 \frac{\phi}{2}\bra{+} U(\bar T', \alpha)\ket{+}\\
\nonumber&-&i\sin\frac{\phi}{2}\cos \frac{\phi}{2}\left(\bra{+} U(\bar T', \alpha)\ket{-}+\bra{-} U(\bar T', \alpha)\ket{+}\right)\\
\nonumber&-&\sin^2\frac{\phi}{2} \bra{-} U(\bar T', \alpha)\ket{-},
\eea
and we now use $\ave{B_\alpha|-}$ is real so
\bea
\nonumber&\cos^2& \frac{\phi}{2}\cos \bar\W \bar T' -2\sin\frac{\phi}{2}\cos \frac{\phi}{2}\sin \bar\W \bar T' \ave{B_\alpha|-}\\
\nonumber&-&\sin^2\frac{\phi}{2} \left(\left|\ave{-|B_\alpha} \right|^{2}\cos \bar\W \bar T'+\left|\ave{-|D_\alpha} \right|^{2}\right)\\
\nonumber&=& -\cos^4 \frac{\phi}{2}-2\cos^2 \frac{\phi}{2}\sin^2 \frac{\phi}{2} -\sin^4\frac{\phi}{2}=-1,
\eea

Summig up, we have demonstrated
\be
U(\bar T', \alpha)=e^{i\alpha}\ket{\phi}\bra{0}+e^{-i\alpha}\ket{0}\bra{-\phi}-\ket{\pi+\phi}\bra{\pi-\phi}.
\ee

\backmatter

\section*{Abbreviations}

\begin{itemize}
\item ERC - Effective Raman coupling.
\item MW - Microwave.
\item NMR - Nuclear magnetic resonance.
\item NV - Nitrogen vacancy.
\end{itemize}

\section*{Declarations}

\begin{itemize}
\item Data and materials availability 

Data sharing is not applicable to this article as no datasets were generated or analysed during the current study.

\item Competing interests 

The authors declare that they have no competing interests.

\item Funding

J.C. acknowledges support from grant PID2021-124965NB-C22 funded by MICIU/AEI/10.13039/501100011033 and by ``ERDF/EU''.
A.L.G. and J.C. acknowledge support from European Union project C-QuENS (Grant No. 101135359). J.C. additionally acknowledges support from grant CNS2023-144994 funded by MICIU/AEI/10.13039/501100011033 and by ``ERDF/EU''.

\item Author contribution

A.L.G.: Writing – review and editing, Numerical calculations,  Figure production, Validation, Investigation, Conceptualization. J.C.: Writing – original draft, Validation, Project administration, Investigation, Formal analysis, Funding acquisition, Conceptualization.

\item Acknowledgments

Not applicable.
\end{itemize}

\end{document}